# A New Algorithm for Inverting General Cyclic Heptadiagonal Matrices Recursively

A.A. KARAWIA[1]
Computer Science Unit, Deanship of Educational Services,
Qassim University, Buraidah 51452, Saudi Arabia.
kraoieh@qu.edu.sa

## ABSTRACT

In this paper, we describe a reliable symbolic computational algorithm for inverting general cyclic heptadiagonal matrices by using parallel computing along with recursion. The algorithm is implementable to the Computer Algebra System(CAS) such as MAPLE, MATLAB and MATHEMATICA. An example is presented for the sake of illustration.

## 1. INTRODUCTION

The n × n general periodic heptadiagonal matrices are takes the form:

$$H = \begin{bmatrix} d_1 & a_1 & A_1 & C_1 & 0 & 0 & \cdots & \cdots & & b_1 \\ b_2 & d_2 & a_2 & A_2 & C_2 & 0 & \cdots & \cdots & & B_2 \\ B_3 & b_3 & d_3 & a_3 & A_3 & C_3 & 0 & \ddots & \vdots & \\ D_3 & B_3 & b_3 & d_3 & a_3 & A_3 & C_3 & 0 & \vdots & \vdots \\ 0 & \ddots & \ddots & \ddots & \ddots & \ddots & \ddots & \ddots & \ddots & \vdots \\ \vdots & \ddots & \ddots & \ddots & \ddots & \ddots & \ddots & \ddots & \ddots & \\ \vdots & \cdots & & D_{n-3} & B_{n-3} & b_{n-3} & d_{n-3} & a_{n-3} & A_{n-3} & C_{n-3} \\ 0 & \cdots & \cdots & & D_{n-2} & B_{n-2} & b_{n-2} & d_{n-2} & a_{n-2} & A_{n-2} \\ A_{n-1} & 0 & \cdots & \cdots & & D_{n-1} & B_{n-1} & b_{n-1} & d_{n-1} & a_{n-1} \\ a_n & A_n & 0 & \cdots & & 0 & D_n & B_n & b_n & d_n \end{bmatrix} \tag{1.1}$$

Where $n \geq 8$.

The inverses of cyclic heptadiagonal matrices are usually required in science and engineering applications, for more details (see special cases, [1-9]). The motivation of the current paper is to establish efficient algorithms for inverting cyclic heptadiagonal matrices of the form (1.1) and for solving linear systems of the form:

$$Hx = r \tag{1.2}$$

[1] Home address: Mathematics Department, Faculty of Science, Mansoura University, Mansoura, 35516, Egypt. E-mail:abibka@mans.edu.eg

where $x=(x_1,x_2,\ldots,x_n)^T$ , $r=(r_1,r_2,\ldots,r_n)^T$ .

To the best of our knowledge, the inversion of a general cyclic heptadiagonal matrix of the form (1.1) has not been considered. Very recently in [9], the inversion of a general cyclic pentadiagonal matrix using recursion is studied without imposing any restrictive conditions on the elements of the matrix. Also, in this paper we are going to compute the inverse of a general cyclic heptadiagonal matrix of the form (1.1) without imposing any restrictive conditions on the elements of the matrix H in (1.1). Our approach is mainly based on getting the elements of the last four columns of $H^{-1}$ in suitable forms via the Doolittle LU factorization [10] along with parallel computation [6]. Then the elements of the remaining (n - 4) columns of $H^{-1}$ may be obtained using relevant recursive relations. The inversion algorithm of this paper is a natural generalization of the algorithm presented in [9]. The development of a symbolic algorithm is considered in order to remove all cases where the numerical algorithm fails.

The paper is organized as follows. In Section 2, new symbolic computational algorithms, that will not break, is constructed. In Section 3, an illustrative example is given. Conclusions of the work are given in Section 4.

## 2. Main results

In this section we shall focus on the construction of new symbolic computational algorithms for computing the determinant and the inverse of general cyclic heptadiagonal matrices. The solution of cyclic heptadiagonal linear systems of the form (1.2) will be taken into account. Firstly we begin with computing the LU factorization of the matrix H. It is as in the following:

$$H=LU \tag{2.1}$$

where

$$L=\begin{bmatrix}
1 & 0 & 0 & 0 & 0 & \cdots & & 0 & 0 & 0\\
f_2 & 1 & 0 & 0 & 0 & \cdots & & 0 & 0 & 0\\
e_3 & f_3 & 1 & 0 & 0 & \cdots & & 0 & 0 & 0\\
\frac{D_4}{\alpha_1} & e_4 & f_4 & 1 & 0 & \cdots & & 0 & 0 & 0\\
0 & \ddots & \ddots & \ddots & \ddots & \ddots & \vdots & \vdots & \vdots & \vdots\\
\vdots & \ddots & \ddots & \ddots & \ddots & \ddots & \ddots & \vdots & \vdots & \vdots\\
0 & 0 & 0 & \ddots & & & & & \vdots & \vdots\\
0 & 0 & 0 & \cdots & \frac{D}{\alpha_{n-5}} & & & & & \vdots\\
k_1 & k_2 & k_3 & \cdots & & k_{n-4} & k_{n-3} & k_{n-2} & 1 & 0\\
h_1 & h_2 & h_3 & & h_{n-5} & h_{n-4} & h_{n-3} & h_{n-2} & h_{n-1} & 1
\end{bmatrix} \tag{2.2}$$

and

$$U=\begin{bmatrix}
\alpha_1 & g_1 & z_1 & C_1 & 0 & \cdots & & 0 & w_1 & v_1\\
0 & \alpha_2 & g_2 & z_2 & \ddots & \ddots & & 0 & w_2 & v_2\\
0 & 0 & \alpha_3 & g_3 & \ddots & \ddots & & 0 & w_3 & v_3\\
\vdots & & \ddots & \ddots & \ddots & \ddots & \vdots & \vdots & \vdots & \vdots\\
\vdots & & & \ddots & \ddots & \ddots & & {}_{n-5} & w_{n-5} & v_{n-5}\\
\vdots & & & & \ddots & \ddots & & & w_{n-4} & v_{n-4}\\
\vdots & & & & & \ddots & & & & v_{n-3}\\
\vdots & & & & & & 0 & \alpha_{n-2} & w_{n-2} & v_{n-2}\\
\vdots & & & & & & 0 & 0 & \alpha_{n-1} & v_{n-1}\\
0 & 0 & 0 & \cdots & \cdots & \cdots & & 0 & 0 & \alpha_n
\end{bmatrix} \qquad (2.3)$$

The elements in the matrices L and U in (2.2) and (2.3) satisfy:

$$\alpha_i=\begin{cases}
d_1 & \text{if } i=1\\
d_2-f_2 g_1 & \text{if } i=2\\
d_3-e_3 z_1-f_3 g_2 & \text{if } i=3\\
d_i-\dfrac{D_i}{\alpha_{i-3}}C_{i-3}-e_i z_{i-2}-f_i g_{i-1} & \text{if } i=4(5)n-2\\
d_{n-1}-\displaystyle\sum_{j=1}^{n-2} w_j k_j & \text{if } i=n-1\\
d_n-\displaystyle\sum_{j=1}^{n-1} v_j h_j & \text{if } i=n,
\end{cases} \qquad (2.4)$$

$$k_i=\begin{cases}
\dfrac{A_{n-1}}{\alpha_1} & \text{if } i=1\\
-\dfrac{k_1 g_1}{\alpha_2} & \text{if } i=2\\
-\dfrac{(k_1 z_1+k_2 g_2)}{\alpha_3} & \text{if } i=3\\
-\dfrac{(k_{i-3}C_{i-3}+k_{i-2}z_{i-2}+k_{i-1}g_{i-1})}{\alpha_i} & \text{if } i=4(5)n-5\\
\dfrac{(D_{n-1}-k_{n-7}C_{n-7}-k_{n-6}z_{n-6}-k_{n-5}g_{n-5})}{\alpha_{n-4}} & \text{if } i=n-4\\
\dfrac{(B_{n-1}-k_{n-6}C_{n-6}-k_{n-5}z_{n-5}-k_{n-4}g_{n-4})}{\alpha_{n-3}} & \text{if } i=n-3\\
\dfrac{(b_{n-1}-k_{n-5}C_{n-5}-k_{n-4}z_{n-4}-k_{n-3}g_{n-3})}{\alpha_{n-2}} & \text{if } i=n-2,
\end{cases} \qquad (2.5)$$

$$h_i = \begin{cases} \dfrac{a_n}{\alpha_1} & \text{if } i = 1 \\ \dfrac{A_n - h_1 g_1}{\alpha_2} & \text{if } i = 2 \\ -\dfrac{(h_1 z_1 + h_2 g_2)}{\alpha_3} & \text{if } i = 3 \\ -\dfrac{(h_{i-3} C_{i-3} + h_{i-2} z_{i-2} + h_{i-1} g_{i-1})}{\alpha_i} & \text{if } i = 4(5)n-4 \\ \dfrac{(D_n - h_{n-6} C_{n-6} - h_{n-5} z_{n-5} - h_{n-4} g_{n-4})}{\alpha_{n-3}} & \text{if } i = n-3 \\ \dfrac{(B_n - h_{n-5} C_{n-5} - h_{n-4} z_{n-4} - h_{n-3} g_{n-3})}{\alpha_{n-2}} & \text{if } i = n-2 \\ \dfrac{(b_n - \sum_{j=1}^{n-2} h_j w_j)}{\alpha_{n-1}} & \text{if } i = n-1, \end{cases} \quad (2.6)$$

$$v_i = \begin{cases} b_1 & \text{if } i = 1 \\ B_2 - f_2 v_1 & \text{if } i = 2 \\ -e_3 v_1 - f_3 v_2 & \text{if } i = 3 \\ -\dfrac{D_i}{\alpha_{i-3}} v_{i-3} - e_i v_{i-2} - f_i v_{i-1} & \text{if } i = 4(5)n-4 \\ C_{n-3} - \dfrac{D_{n-3}}{\alpha_{n-6}} v_{n-6} - e_{n-3} v_{n-5} - f_{n-3} v_{n-4} & \text{if } i = n-3 \\ A_{n-2} - \dfrac{D_{n-2}}{\alpha_{n-5}} v_{n-5} - e_{n-2} v_{n-4} - f_{n-2} v_{n-3} & \text{if } i = n-2 \\ a_{n-1} - \sum_{j=1}^{n-2} v_j k_j & \text{if } i = n-1, \end{cases} \quad (2.7)$$

$$w_i = \begin{cases} B_1 & \text{if } i = 1 \\ -f_2 w_1 & \text{if } i = 2 \\ -f_3 w_2 - e_3 w_1 & \text{if } i = 3 \\ -\dfrac{D_i}{\alpha_{i-3}} w_{i-3} - e_i w_{i-2} - f_i w_{i-1} & \text{if } i = 4(5)n-5 \\ C_{n-4} - \dfrac{D_{n-4}}{\alpha_{n-7}} W_{n-7} - e_{n-4} w_{n-6} - f_{n-4} w_{n-5} & \text{if } i = n-4 \\ A_{n-3} - \dfrac{D_{n-3}}{\alpha_{n-6}} w_{n-6} - e_{n-3} w_{n-5} - f_{n-3} w_{n-4} & \text{if } i = n-3 \\ a_{n-2} - \dfrac{D_{n-2}}{\alpha_{n-5}} w_{n-5} - e_{n-2} w_{n-4} - f_{n-2} w_{n-3} & \text{if } i = n-2, \end{cases} \quad (2.8)$$

$$f_i = \begin{cases} \dfrac{b_2}{\alpha_1} & \text{if } i = 2 \\ \dfrac{b_3 - e_3 g_1}{\alpha_2} & \text{if } i = 3 \\ \dfrac{b_i - \dfrac{D_i}{\alpha_{i-3}} z_{i-3} - e_i g_{i-2}}{\alpha_{i-1}} & \text{if } i = 4(5)n-2, \end{cases} \tag{2.9}$$

$$e_i = \begin{cases} \dfrac{B_3}{\alpha_1} & \text{if } i = 3 \\ \dfrac{B_i - \dfrac{D_i}{\alpha_{i-3}} g_{i-3}}{\alpha_{i-2}} & \text{if } i = 4(5)n-2, \end{cases} \tag{2.10}$$

$$g_i = \begin{cases} a_1 & \text{if } i = 1 \\ a_2 - f_2 z_1 & \text{if } i = 2 \\ a_i - f_i z_{i-1} - e_i C_{i-2} & \text{if } i = 3(4)n-3, \end{cases} \tag{2.11}$$

$$z_i = \begin{cases} A_1 & \text{if } i = 1 \\ A_i - f_i C_{i-1} & \text{if } i = 2(3)n-4. \end{cases} \tag{2.12}$$

We also have:

$$\det H = \prod_{i=1}^{n} \alpha_i . \tag{2.13}$$

At this point it is convenient to formulate our first result. It is a symbolic algorithm for computing the determinant of a cyclic heptadiagonal matrix H of the form (1.1) and can be considered as natural generalization of the symbolic algorithm **DETCPENTA** in [9].

**Algorithm 2.1.** To compute det H for the cyclic heptadiagonal matrix H in (1.1), we may proceed as follows:

**Step 1:** Set $\alpha_1$=$d_1$. If $\alpha_1$=0 then $\alpha_1$= t end if. Set $g_1$=$a_1$, $z_1$=$A_1$, $k_1$=$A_{n-1}$/ $\alpha_1$, $v_1$=$b_1$, $w_1$=B1, $h_1$=$a_n$/ $\alpha_1$, $w_1$=$B_1$ , $f_2$=$b_2$/ $\alpha_1$, $e_3$=$B_3$/ $\alpha_1$, $\alpha_2$=$d_2$-$f_2$ *$g_1$. If $\alpha_2$=0 then $\alpha_2$= t end if. Set $K_2$=-$k_1$*$g_1$/ $\alpha_2$, $v_2$=$B_2$-$f_2$*$v_1$, $w_2$=-$f_2$*$w_1$, $h_2$=($A_n$-$h_1$*$g_1$)/ $\alpha_2$, $g_2$=$a_2$-$f_2$*$z_1$, $f_3$=($b_3$-$e_3$*$g_1$)/$\alpha_2$, $\alpha_3$=$d_3$-$e_3$*$z_1$-$f_3$*$g_2$. If $\alpha_3$=0 then $\alpha_3$= t end if. Set $k_3$=-($k_1$*$z_1$+$k_2$*$g_2$)/$\alpha_3$, $h_3$=-($h_1z_1$+$h_2g_2$)/$\alpha_3$, $v_3$=-$e_3$*$v_1$-$f_3$*$v_2$, $w_3$=-$f_3w_2$-$e_3w_1$.

**Step 2:** Compute and simplify:

For i from 4 to n-2 do
- $e_i$=($B_i$-$D_i$* $g_{i-3}$/ $\alpha_{i-3}$)/ $\alpha_{i-2}$
- $f_i$=($b_i$- $D_i$* $z_{i-3}$/ $\alpha_{i-3}$-$e_i$*$g_{i-2}$)/ $\alpha_{i-1}$
- $z_{i-2}$=$A_{i-2}$-$f_{i-2}$*$C_{i-3}$
- $g_{i-1}$=$a_{i-1}$-$f_{i-1}$*$z_{i-2}$-$e_{i-1}$*$C_{i-3}$
- $\alpha_i$=($d_i$- $D_i$* $C_{i-3}$/ $\alpha_{i-3}$-$e_i$*$z_{i-2}$-$f_i$*$g_{i-1}$)
- If $\alpha_i$=0 then $\alpha_i$= t end if

End do

**Step 3:** Compute and simplify:

For i from 4 to n-5 do
- $K_i$=-($k_{i-3}$*$C_{i-3}$+$k_{i-2}$*$z_{i-2}$+$k_{i-1}$*$g_{i-1}$)/ $\alpha_i$
- $w_i$=-($D_i$*$w_{i-3}$/ $\alpha_{i-3}$+$e_i$*$w_{i-2}$+$f_i$*$w_{i-1}$)

End do

**Step 4:** Compute and simplify:

For i from 4 to n-4 do

$h_i=-(h_{i-3}*C_{i-3}+h_{i-2}*z_{i-2}+h_{i-1}*g_{i-1})/\alpha_i$

$v_i=-(D_i*v_{i-3}/\alpha_{i-3}+e_i*v_{i-2}+f_i*v_{i-1})$

End do

**Step 5:** Compute simplify:

$k_{n-4}=(D_{n-1}-k_{n-5}*g_{n-5}-k_{n-6}*z_{n-6}-k_{n-7}*C_{n-7})/\alpha_{n-4}$

$k_{n-3}=(B_{n-1}-k_{n-4}*g_{n-4}-k_{n-5}*z_{n-5}-k_{n-6}*C_{n-6})/\alpha_{n-3}$

$k_{n-2}=(b_{n-1}-k_{n-3}*g_{n-3}-k_{n-4}*z_{n-4}-k_{n-5}*C_{n-5})/\alpha_{n-2}$

$w_{n-4}=C_{n-4}-D_{n-4}*w_{n-7}/\alpha_{n-7}-e_{n-4}*w_{n-6}-f_{n-4}w_{n-5}$

$w_{n-3}=A_{n-3}-D_{n-3}*w_{n-6}/\alpha_{n-6}-e_{n-3}*w_{n-5}-f_{n-3}w_{n-4}$

$w_{n-2}=a_{n-2}-D_{n-2}*w_{n-5}/\alpha_{n-5}-e_{n-2}*w_{n-4}-f_{n-2}w_{n-3}$

$h_{n-3}=(D_n-h_{n-4}*g_{n-4}-h_{n-5}*z_{n-5}-h_{n-6}*C_{n-6})/\alpha_{n-3}$

$h_{n-2}=(B_n-h_{n-3}*g_{n-3}-h_{n-4}*z_{n-4}-h_{n-5}*C_{n-5})/\alpha_{n-2}$

$v_{n-3}=C_{n-3}-D_{n-3}*v_{n-6}/\alpha_{n-6}-e_{n-3}*v_{n-5}-f_{n-3}*v_{n-4}$

$v_{n-2}=A_{n-2}-D_{n-2}*v_{n-5}/\alpha_{n-5}-e_{n-2}*v_{n-4}-f_{n-2}*v_{n-3}$

$v_{n-1}=a_{n-1}-\sum_{j=1}^{n-2}k_jv_j$

$\alpha_{n-1}=d_{n-1}-\sum_{j=1}^{n-2}k_jw_j$

If $\alpha_{n-1}=0$ then $\alpha_{n-1}=t$ end if

$h_{n-1}=(b_n-\sum_{j=1}^{n-2}h_jw_j)/\alpha_{n-1}$

$\alpha_n=d_n-\sum_{j=1}^{n-1}h_jv_j$

If $\alpha_n=0$ then $\alpha_n=t$ end if

**Step 6:** Compute det H = $\left(\prod_{i=1}^{n}\alpha_i\right)_{t=0}$.

The symbolic Algorithm 2.1 will be referred to as **DETCHEPTA**. The new algorithm **DETCHEPTA** is very useful to check the nonsingularity of the matrix H when we consider, for example, the solution of the cyclic heptadiagonal linear systems of the form (1.2).

Now, when the matrix H is nonsingular, its inversion is computed as follows:

Let $$H^{-1}=(S_{ij})_{1\le i,j\le n}=(Col_1,Col_2,...,Col_n) \quad (2.14)$$

Where $Col_m$ denotes the m[th] column of $H^{-1}$, $m=1,2,...,n$.

Since the Doolittle LU factorization of the matrix H in (1.1) is always possible then we can use parallel computations to get the elements of the last five columns $Col_i=(S_{1,i},S_{2,i},...,S_{n,i})$, $i=n,n-1,n-2,n-3$ and $n-4$ of $H^{-1}$ as follows [9]:

$$S_{n,n}=\frac{1}{\alpha_n}, \quad (2.15)$$

$$S_{n-1,n}=-\frac{v_{n-1}S_{n,n}}{\alpha_{n-1}}, \tag{2.16}$$

$$S_{i,n}=-\frac{1}{\alpha_i}(g_iS_{i+1,n}+z_iS_{i+2,n}+C_iS_{i+3,n}+w_iS_{n-1,n}+v_iS_{n,n}), i=n-5,n-6,...,1 \tag{2.17}$$

$$S_{n,n-1}=-\frac{h_{n-1}}{\alpha_n}, \tag{2.18}$$

$$S_{n-1,n-1}=\frac{1}{\alpha_{n-1}}(1-v_{n-1}S_{n,n-1}), \tag{2.19}$$

$$S_{i,n-1}=-\frac{1}{\alpha_i}(g_iS_{i+1,n-1}+z_iS_{i+2,n-1}+C_iS_{i+3,n-1}+w_iS_{n-1,n-1}+v_iS_{n,n-1}), i=n-5,n-4,...,1 \tag{2.20}$$

$$S_{n,n-2}=\frac{-h_{n-2}+h_{n-1}k_{n-2}}{\alpha_n}, \tag{2.21}$$

$$S_{n-1,n-2}=-\frac{1}{\alpha_{n-1}}(k_{n-2}+v_{n-1}S_{n,n-2}), \tag{2.22}$$

$$S_{n-2,n-2}=\frac{1}{\alpha_{n-2}}(1-w_{n-2}S_{n-1,n-2}-v_{n-2}S_{n,n-2}), \tag{2.23}$$

$$S_{i,n-2}=-\frac{1}{\alpha_i}(g_iS_{i+1,n-2}+z_iS_{i+2,n-2}+C_iS_{i+3,n-2}+w_iS_{n-1,n-2}+v_iS_{n,n-2}), i=n-5,n-6,...,1 \tag{2.24}$$

$$S_{n,n-3}=\frac{-h_{n-3}+h_{n-2}f_{n-2}-h_{n-1}(k_{n-2}f_{n-2}-k_{n-3})}{\alpha_n}, \tag{2.25}$$

$$S_{n-1,n-3}=\frac{1}{\alpha_{n-1}}(k_{n-2}f_{n-2}-k_{n-3}-v_{n-1}S_{n,n-3}), \tag{2.26}$$

$$S_{n-2,n-3}=-\frac{1}{\alpha_{n-2}}(f_{n-2}+w_{n-2}S_{n-1,n-3}+v_{n-2}S_{n,n-3}), \tag{2.27}$$

$$S_{n-3,n-3}=\frac{1}{\alpha_{n-3}}(1-g_{n-3}S_{n-2,n-3}-w_{n-3}S_{n-1,n-3}-v_{n-3}S_{n,n-3}), \tag{2.28}$$

$$S_{n,n-4}=\frac{1}{\alpha_n}(-h_{n-4}+h_{n-3}f_{n-3}+h_{n-2}(e_{n-2}-f_{n-2}f_{n-3})+h_{n-1}(k_{n-4}-k_{n-3}f_{n-3}-k_{n-2}(e_{n-2}+f_{n-2}f_{n-3}))) \tag{2.29}$$

$$S_{n-1,n-4}=\frac{1}{\alpha_{n-1}}(-k_{n-4}+k_{n-3}f_{n-3}+k_{n-2}(e_{n-2}-f_{n-2}f_{n-3})-v_{n-1}S_{n,n-4}) \tag{2.30}$$

$$S_{n-2,n-4}=-\frac{1}{\alpha_{n-2}}((e_{n-2}-f_{n-2}f_{n-3})+w_{n-2}S_{n-1,n-4}+v_{n-2}S_{n,n-4}), \tag{2.31}$$

$$S_{n-3,n-4}=-\frac{1}{\alpha_{n-3}}(f_{n-3}+g_{n-3}S_{n-2,n-4}+w_{n-3}S_{n-1,n-4}+v_{n-3}S_{n,n-4}), \tag{2.32}$$

$$S_{n-4,n-4}=\frac{1}{\alpha_{n-4}}(1-g_{n-4}S_{n-3,n-4}-z_{n-4}S_{n-2,n-4}-w_{n-4}S_{n-1,n-4}-v_{n-4}S_{n,n-4}), \tag{2.33}$$

$$S_{i,n-4} = \frac{-1}{\alpha_i}(g_i S_{i+1,n-4} + z_i S_{i+2,n-4} + C_i S_{i+3,n-4} + w_i S_{n-1,n-4} + v_i S_{n,n-4}),\ i = n-5, n-6, ..., 1 \quad (2.34)$$

$$S_{n-2,j} = -\frac{1}{\alpha_{n-2}}(w_{n-2} S_{n-1,j} + v_{n-2} S_{n,j}),\ j = n, n-1 \quad (2.35)$$

$$S_{n-3,j} = -\frac{1}{\alpha_{n-3}}(g_{n-3} S_{n-2,j} + w_{n-3} S_{n-1,j} + v_{n-3} S_{n,j}),\ j = n, n-1, n-2 \quad (2.36)$$

$$S_{n-4,j} = -\frac{1}{\alpha_{n-4}}(g_{n-4} S_{n-3,j} + z_{n-4} S_{n-2,j} + w_{n-4} S_{n-1,j} + v_{n-4} S_{n,j}), j = n, n-1, n-2, n-3 \quad (2.37)$$

$$S_{i,j} = -\frac{1}{\alpha_i}(g_i S_{i+1,j} + A_i S_{i+2,j} + w_i S_{n-1,j} + v_i S_{n,j}), i = n-5, n-6, ..., 1,\ j = n, n-1, ..., n-4 \quad (2.38)$$

The remaining (n - 5) columns are obtained by using the fact $H^{-1}H = I_n$, where $I_n$ is the identity matrix. They are as in the following:

$$Col_{n-5} = \frac{1}{C_{n-5}}(E_{n-2} - A_{n-4}Col_{n-4} - a_{n-3}Col_{n-3} - d_{n-2}Col_{n-2} - b_{n-1}Col_{n-1} - B_n Col_n), \quad (2.39)$$

$$Col_j = \frac{1}{C_j}(E_{j+3} - A_{j+1}Col_{j+1} - a_{j+2}Col_{j+2} - d_{j+3}Col_{j+3} - b_{j+4}Col_{j+4} - B_{j+5}Col_{j+5} - D_{j+6}Col_{j+6}).$$
$$j = n-6, n-7, ..., 1 \quad (2.40)$$

where $E_i$=1 if row=i.

**Remark 2.1.** Eqs. (2.39) and (2.40) suggest an additional assumption $\prod_{i=1}^{n-5} C_i \neq 0,$, which is only formal and can be omitted by introducing auxiliary parameter t in Algorithm 2.2 given below.

Now we formulate a second result. It is a symbolic computational algorithm to compute the inverse of a general cyclic heptadiagonal matrix of the form (1.1) when it exists.

**Algorithm 2.2.** To find the n x n inverse matrix of the general cyclic heptadiagonal matrix H in (1.1) by using the relations (2.15)_(2.40).

**INPUT**: Order of the matrix n and the components, $D_i$, $B_i$, $b_i$, $d_i$, $a_i$, $A_i$, $C_i$, i=1,2, ...,n, where $D_1 = D_2 = D_3 = C_{n-2} = C_{n-1} = C_n = 0$.

**OUTPUT:** Inverse matrix, $H^{-1} = (S_{ij})_{1 \le i,j \le n}$.

**Step 1:** If $C_i$=0 for any i = 1, 2, ..., n-5 set $C_i$= t (t is just a symbolic name).

**Step 2:** If $B_i$=0 for any i = 6, 7, ..., n, $B_i$=t.

**Step 3:** Use the **DETCHEPTA** algorithm to check the nonsingularity of the matrix H. If the matrix H is singular then OUTPUT (`The matrix H is singular'); Stop.

**Step 4:** For i = 1, 2, ..., n, compute and simplify the components $S_{i,n}$, $S_{i,n-1}$, $S_{i,n-2}$, $S_{i,n-3}$ and $S_{i,n-4}$ of the columns $C_j$, j=n, n-1, n-2, n-3 and n-4, respectively, by using (2.15)-(2.38).

**Step 5:** For i = 1, 2, ..., n, compute and simplify the components $S_{i,n-5}$ by using (2.39).

**Step 6:** For j = n-6, n-7, ..., 1, do

For i = 1, 2, ..., n, do

Compute and simplify the components $S_{i,j}$ by using (2.40).

End do

End do.

**Step 6:** Substitute the actual value t = 0 in all expressions to obtain the elements, $S_{i,j}$, i, j = 1, 2, ..., n.

The symbolic **Algorithm 2.2** will be referred to as **CHINV** algorithm. The Algorithms 2.2, 2.3 and 2.2 in [9], [11] and [12], respectively, are now special cases of the **CHINV** algorithm.

## 3. An illustrative example

In this section we give an example for the sake of illustration.

**Example 3.1.** Consider the 10 X 10 cyclic heptadiagonal linear system

$$\begin{bmatrix} 1 & -1 & 1 & -2 & 0 & 0 & 0 & 0 & 2 & -1 \\ 1 & 1 & 1 & 1 & -1 & 0 & 0 & 0 & 0 & 1 \\ 2 & 1 & -1 & 1 & 2 & 3 & 0 & 0 & 0 & 0 \\ 2 & -2 & 3 & 1 & 5 & -6 & 0 & 0 & 0 & 0 \\ 0 & 1 & 1 & 1 & 1 & 1 & 1 & 2 & 0 & 0 \\ 0 & 0 & -1 & -1 & -1 & -1 & -1 & -1 & 1 & 0 \\ 0 & 0 & 0 & 2 & 2 & 2 & 2 & 3 & 1 & -3 \\ 0 & 0 & 0 & 0 & -2 & -2 & 1 & 1 & 3 & 5 \\ 3 & 0 & 0 & 0 & 0 & 3 & 1 & 3 & 4 & -1 \\ 2 & 4 & 0 & 0 & 0 & 0 & 2 & 3 & 4 & 1 \end{bmatrix} \begin{bmatrix} x_1 \\ x_2 \\ x_3 \\ x_4 \\ x_5 \\ x_6 \\ x_7 \\ x_8 \\ x_9 \\ x_{10} \end{bmatrix} = \begin{bmatrix} 2 \\ 15 \\ 33 \\ 0 \\ 43 \\ -24 \\ 47 \\ 70 \\ 78 \\ 94 \end{bmatrix} \qquad (3.1)$$

By using the coefficient matrix of the system (3.1) and applying the **CHINV** algorithm, we get:

$$H^{-1} = \begin{bmatrix}
-\frac{12664}{32715} & -\frac{2921}{32715} & -\frac{1898}{10905} & \frac{2957}{32715} & -\frac{23399}{32715} & -\frac{24419}{32715} & -\frac{2069}{6543} & -\frac{6676}{32715} & \frac{13714}{32715} & \frac{6316}{32715} \\
\frac{2686}{32715} & \frac{4169}{32715} & \frac{902}{10905} & -\frac{1118}{32715} & \frac{8006}{32715} & \frac{8801}{32715} & \frac{41}{6543} & -\frac{2366}{32715} & -\frac{6391}{32715} & \frac{4571}{32715} \\
\frac{5417}{10905} & \frac{4693}{10905} & \frac{344}{3635} & -\frac{241}{10905} & \frac{5842}{10905} & \frac{3292}{10905} & \frac{280}{2181} & \frac{758}{10905} & -\frac{2792}{10905} & -\frac{1658}{10905} \\
\frac{293}{10905} & \frac{5347}{10905} & \frac{591}{3635} & \frac{146}{10905} & \frac{1393}{10905} & \frac{5698}{10905} & \frac{853}{2181} & \frac{1577}{10905} & -\frac{1838}{10905} & -\frac{1982}{10905} \\
\frac{6344}{32715} & -\frac{4964}{32715} & \frac{2983}{10905} & \frac{1598}{32715} & \frac{12259}{32715} & \frac{9484}{32715} & \frac{433}{6543} & \frac{2621}{32715} & -\frac{6374}{32715} & -\frac{1676}{32715} \\
\frac{938}{3635} & \frac{357}{3635} & \frac{788}{3635} & -\frac{339}{3635} & \frac{1023}{3635} & \frac{513}{3635} & \frac{56}{727} & \frac{297}{3635} & -\frac{413}{3635} & -\frac{477}{3635} \\
\frac{1178}{10905} & -\frac{908}{10905} & -\frac{254}{3635} & -\frac{604}{10905} & -\frac{13232}{10905} & -\frac{14012}{10905} & \frac{385}{2181} & \frac{497}{10905} & -\frac{1658}{10905} & \frac{3718}{10905} \\
-\frac{6356}{10905} & -\frac{4969}{10905} & -\frac{1382}{3635} & \frac{778}{10905} & \frac{3539}{10905} & -\frac{1306}{10905} & -\frac{922}{2181} & -\frac{1904}{10905} & \frac{5891}{10905} & \frac{194}{10905} \\
\frac{16382}{32715} & \frac{10738}{32715} & \frac{3244}{10905} & -\frac{1216}{32715} & \frac{14092}{32715} & \frac{27832}{32715} & \frac{2725}{6543} & \frac{8078}{32715} & -\frac{11282}{32715} & -\frac{5153}{32715} \\
-\frac{808}{32715} & -\frac{3617}{32715} & \frac{1174}{10905} & \frac{44}{32715} & \frac{5947}{32715} & -\frac{1868}{32715} & -\frac{938}{6543} & \frac{4658}{32715} & \frac{193}{32715} & -\frac{1643}{32715}
\end{bmatrix}$$

By simple calculation we obtain the solution of cyclic heptadiagonal linear system (3.1) $X=[x_1, x_2, x_3, x_4, x_5, x_6, x_7, x_8, x_9, x_{10}]=[1,2,3,4,5,6,7,8,9,10]$.

## 4. Conclusions

In this work new recursive computational algorithms have been developed for computing the determinant and inverse of general cyclic heptadiagonal matrices and for solving linear systems of cyclic heptadiagonal type. The algorithms are reliable, computationally efficient and will not fail. The algorithms are natural generalizations of some algorithms in current use.

## References

[1] A.A. Karawia, A computational algorithm for solving periodic pentadiagonal linear systems, Appl. Math. Comput. 174 (2006) 613_618.

[2] M. Batista, A method for solving cyclic block penta-diagonal systems of linear equations, arXiv:0806.3639V5 [math-ph].

[3] I.M. Navon, A periodic pentadiagonal systems solver, Commun. Appl. Numer. Methods 3 (1987) 63_69.

[4] X.-G. Lv, J. Le, A note on solving nearly pentadiagonal linear systems, Appl. Math. Comput. 204 (2008) 707_712.

[5] S.N. Neossi Nguetchue, S. Abelman, A computational algorithm for solving nearly penta-diagonal linear systems, Appl. Math. Comput. 203 (2008) 629_634.

[6] T. Sogabe, New algorithms for solving periodic tridiagonal and periodic pentadiagonal linear systems, Appl. Math. Comput. 202 (2008) 850_856.

[7] X.i-Le Zhao, Ting-Zhu Huang, On the inverse of a general pentadiagonal matrix, Appl. Math. Comput. 202 (2008) 639_646.

[8] A. Driss Aiat Hadj, M. Elouafi, A fast numerical algorithm for the inverse of a tridiagonal and pentadiagonal matrix, Appl. Math. Comput. 202 (2008) 441_445.

[9] M. El-Mikkawy, El-Desouky Rahmo, Symbolic algorithm for inverting cyclic pentadiagonal matrices recursively - Derivation and implementation, Computers and Mathematics with Applications 59 (2010) 1386-1396.
[10] J.W. Demmel, Numerical linear algebra, SIAM (1997).
[11] M. El-Mikkawy, E. Rahmo, A new recursive algorithm for inverting general tridiagonal and anti-tridiagonal matrices, Appl. Math. Comput. 204 (2008) 368_372.
[12] M. El-Mikkawy, E. Rahmo, A new recursive algorithm for inverting general periodic pentadiagonal and anti-pentadiagonal matrices, Appl. Math. Comput. 207 (2009) 164_170.